\documentstyle[amssymb,preprint,aps]{revtex}
\tightenlines 

\begin{document}
\title{Crystallization of the ordered vortex phase in high temperature
superconductors}
\author{D. Giller, B. Ya. Shapiro, I. Shapiro, A. Shaulov, and Y. Yeshurun}
\address{\ \\
{\it Institute of Superconductivity, Department of Physics,} {\it Bar-Ilan}\\
University, {\it Ramat Gan 5290, Israel}}
\maketitle

\begin{abstract}
The Landau-Khalatnikov time-dependent equation is applied to describe the
crystallization process of the ordered vortex lattice in high temperature
superconductors after a sudden application of a magnetic field. Dynamic
coexistence of a stable ordered phase and an unstable disordered phase, with
a sharp interface between them, is demonstrated. The transformation to the
equilibrium ordered state proceeds by movement of this interface from the
sample center toward its edge. The theoretical analysis dictates specific
conditions for the creation of a propagating interface, and provides the
time scale for this process.

\end{abstract}

The process of formation of various equilibrium phases after a sudden change
in the thermodynamic conditions, is a topic of wide theoretical \cite
{vanSaarloos} and experimental \cite{Cladis}\ interest. Obviously, the
initial state created immediately after the abrupt change is a
non-equilibrium, unstable state. The transformation of this state to the
thermodynamic equilibrium state may proceed either homogeneously throughout
the entire system, or by nucleation of a spatially localized domain of the
equilibrium phase, creating a front which propagates until equilibrium is
reached in the entire system. The latter process has been recently observed
in the vortex system of Bi$_{2}$Sr$_{2}$CaCu$_{2}$O$_{8+\delta }$ (BSCCO)
crystals. \cite{GillerPRL,vanderBeek} This vortex system exhibits a
transition between an ordered and disordered phases, at a
temperature-independent transition field $B_{od}\simeq 400$ $G$.\cite
{Cubitt,muSr} High temporal resolution magneto-optical measurements \cite
{GillerPRL,vanderBeek} indicated that immediately after a sudden application
of external magnetic field $B_{a}\lesssim B_{od}$, a transient disordered
vortex state is created, followed by a nucleation and front propagation of
an ordered vortex state.\cite{GillerPRL} The purpose of this paper is to
analyze theoretically the crystallization process of the vortex ordered
phase, i.e. the nucleation process, the creation of a front and its motion.

Our analysis is based on the Landau-Khalatnikov (LK) time dependent equation 
\cite{Khalatnikov}: 
\begin{equation}
\frac{\partial \Psi }{\partial t}=-\Gamma \frac{\delta F}{\delta \Psi },
\label{GL}
\end{equation}
where $\Psi $ and $F$ are the order parameter and the free energy of the
system, and $\Gamma $ is the Landau-Khalatnikov damping coefficient. We
define the order parameter of the vortex system in a way analogous to the
definition of the order parameter in order-disorder transitions in atomic
solids. \cite{JeballeWhite} In \ the latter case, the order parameter $\rho
_{q}$ is a set of Fourier components of the atomic density taken at
reciprocal lattice vectors $q=G$. In particular, for an ordered lattice
phase $\rho _{q}=const\neq 0$ at $q=G$, whereas for a disordered state $\rho
_{q}=0$ for all $q\neq 0$. Extending this approach to the vortex
order-disorder phase transition, we note that in small-angle neutron
scattering experiments in $BSCCO$, \cite{Cubitt} Bragg peaks are observed at
low temperatures and low fields mainly in the first Brillouin zone; these
peaks are smeared for fields larger than $B_{od}$. Thus, only one component, 
$\rho _{G_{1}}$, is sufficient to completely describe the order parameter, $%
\rho _{G_{1}}$ being the value of the Fourier component of the {\it vortex}
density at the minimal vector of the reciprocal lattice. In order to
describe the kinetics of the phase transition we allow the order parameter
to be temporally and spatially dependent, $\Psi ({\bf r},t)\equiv \rho
_{G_{1}}({\bf r},t)$, assuming that $\Psi ({\bf r})$ varies slowly over the
inter-vortex distance. The scalar real order parameter $\Psi ({\bf r},t)$ so
defined, distinguishes between two thermodynamic solid phases of the vortex
matter: $\Psi =0$ for the disordered state and $\Psi =\Psi _{0}\neq 0$ for
the ordered state.

In the Ginzburg-Landau formalism, the phase transition between the ordered
and disordered phases may be described by a free energy density functional $%
F $: 
\begin{equation}
F=\frac{1}{2}D(\nabla \Psi )^{2}-\frac{1}{2}\alpha \Psi ^{2}-\frac{1}{3}%
\beta \Psi ^{3}+\frac{1}{4}\gamma \Psi ^{4}  \label{F}
\end{equation}
where $\alpha $,$\beta $, $\gamma $ and $D$ are the Landau coefficients.
These coefficients depend on the vortex-vortex and vortex-pinning
interactions, and their evaluation requires a microscopic theory which does
not yet exist. Note that Eq. 2 does not describe the whole free energy of
the vortex system, but only that part which is varying through the phase
transition, i.e. $\Psi -$dependent.

As the order-disorder vortex phase transition in BSCCO is field driven, we
express the parameter $\alpha $ as a function of $B$: 
\begin{equation}
\alpha =\alpha _{0}(1-B/B^{\ast }),  \label{alpha}
\end{equation}
where $B^{\ast }$ is a characteristic field related to the transition field $%
B_{od}=B^{\ast }(1+2/9\mu )$, where $\mu =\alpha _{0}\gamma /\beta ^{2}$.
Note that for a second order transition ($\beta =0$), $B_{od}=B^{\ast }$.
For a first order phase transition, metastable states of the system are
found between $B^{\ast }$ and $B^{\ast \ast }=B^{\ast }(1+1/4\mu )$. For $%
B<B^{\ast }$ the disordered state is unstable while the ordered state,
characterized by 
\[
\Psi =\Psi _{0}=\frac{\beta }{2\gamma }\left( 1+\sqrt{1+4\mu (1-B/B^{\ast })}%
\right) , 
\]
is stable. For $B>B^{\ast \ast }$ the ordered state is unstable, while the
disordered state with $\Psi =0$ is thermodynamically favorable. All the
above results are deduced from the conventional Landau theory for phase
transitions \cite{ChaikinLubensky} by replacing temperature with the
induction $B$.

In solving Eq. (\ref{GL})\ \ we assume an initial non-equilibrium disordered
vortex state ($\Psi =0$) caused by the rapid injection of the vortices
through non-uniform surface barriers. \cite{GillerPRL,Paltiel} We show that
Eq. (\ref{GL}) can describe the nucleation and growth of the vortex ordered
state ($\Psi =\Psi _{0}$). To demonstrate this point we assume an induction
distribution with a constant gradient \cite{Bean} $\widetilde{B}/d$, i.e. $%
B=B_{a}-$ $\widetilde{B}(1-|x|/d)$, where $B_{a}$ is an applied field and $d$
is half-width of the sample. In this case, Eq. (\ref{GL}) can be solved
analytically for both the nucleation and growth processes.

A solution for the nucleation process, i.e. the {\it initial} growth of the
order parameter $\Psi $ ($\Psi $ close to zero), is obtained by neglecting
nonlinear terms in Eq. (\ref{GL}): 
\begin{equation}
\frac{1}{\Gamma }\frac{\partial \Psi }{\partial t}=D\frac{\partial ^{2}\Psi 
}{\partial x^{2}}+\left( \alpha _{0}-\alpha _{0}\frac{B_{a}-\widetilde{B}}{%
B^{\ast }}-\alpha _{0}\frac{\widetilde{B}}{B^{\ast }}\frac{x}{d}\right) \Psi
.  \label{GLlin}
\end{equation}
The boundary conditions dictated by symmetry is $d\Psi (x,t)/dx|_{x=0}=0$;
also we require that $\Psi (x,t)$ is a non-diverging function. The solution
of Eq. (\ref{GLlin}) is then: 
\begin{equation}
\Psi (x,t)=\sum\limits_{n=0}^{\infty }A_{n}e^{\Lambda
_{n}t}Ai(x/x_{s}-\varsigma _{n}),  \label{Psin}
\end{equation}
where 
\begin{equation}
\Lambda _{n}=\Gamma \alpha _{0}\left[ 1-\frac{B_{a}-\widetilde{B}}{B^{\ast }}%
-\varsigma _{n}\left( \frac{a_{D}}{\mu }\right) ^{1/3}\left( \frac{%
\widetilde{B}}{B^{\ast }}\right) ^{2/3}\right] .  \label{lambdan}
\end{equation}
$Ai$ is the Airy function, $\varsigma _{n}=0.685,3.9,7.06,...$ are the
solutions of $J_{2/3}(\varsigma _{n})=J_{-2/3}(\varsigma _{n})$ where $%
J_{\nu }$ is the Bessel function, and $x_{s}=\left( DdB^{\ast }/\alpha _{0}%
\widetilde{B}\right) ^{1/3}=d\left( a_{D}B^{\ast }/\mu \widetilde{B}\right)
^{1/3}$. Here $a_{D}=D\gamma /\beta ^{2}d^{2}$ is a dimensionless{\LARGE \ }%
exchange coefficient. Note that $\varsigma _{n}$ is a constant of order $n\ $%
, growing with increasing $n$.

It is evident from Eq. (\ref{Psin}) that only terms with $\Lambda _{n}>0$\
play a role in the nucleation process. For $B_{a}-\widetilde{B}>B^{\ast },$
i.e. the entire sample is in a metastable or a stable state, all $\Lambda
_{n}$ are negative, implying that the nucleation process cannot take place.
For $B_{a}=\widetilde{B}$ the induction at the center of the sample is zero
and the rate of the nucleation process is maximum. Relation (\ref{lambdan})
shows that the exponent with $n=0$ yields the fastest nucleation rate, thus
governing the nucleation process. This process may thus be approximately
described by the first term in Eq. (\ref{Psin}). \ In this approximation,
the development of the order parameter during the nucleation process is
described by the dashed lines in Fig. 1. Note that the analytical solution (%
\ref{Psin}) describes only the first stages of the nucleation process, where
the non-linear terms in Eq. (\ref{GL}) may be neglected. This solution
ceases to apply when the value of $\Psi $ approaches $\Psi _{0}$, i.e. after
a time period of order $1/\Lambda _{0}.$ The width of the ordered domain is
then $w\sim x_{s}\left( 1+\varsigma _{0}\right) \sim x_{s}.$ The condition
for appearance of {\em localized} domain in the sample center may be then
obtained from the inequality $x_{s}<<d$ or: 
\begin{equation}
\frac{\widetilde{B}}{B^{\ast }}>>\frac{a_{D}}{\mu }.  \label{Condxsd}
\end{equation}
If this condition is not satisfied, then {\em homogeneous} transformation of
the unstable phase takes place. Otherwise, a sharp {\em front} will develop
separating between the nucleating ordered phase and the initial unstable
disordered phase, as described above. Thus, when the induction gradient is
large enough, we expect the appearance of a sharp interface between the
growing stable (ordered) phase and the retreating unstable (disordered)
phase.

In describing the {\it growth} process, i.e. the movement of the interface
between the ordered and disordered phases, non linear terms in Eq. \ref{GL}
must be taken into account.\ We express the linearly varying function $B(x)$
as: $B=B_{f}+(\widetilde{B}/d)(x-x_{f})$, where $B_{f}=(B_{a}-\widetilde{B}%
)+x_{f}\widetilde{B}/d$ is the induction at the front located at $x_{f}$.\ \
In this notation: $\alpha =1-[B_{f}+(\widetilde{B}/d)(x-x_{f})]/B^{\ast }$.
Eq. (\ref{GL}) can be written in the reference frame of an observer moving
with the front, by introducing a new variable $\xi =x-x_{f\text{ }}(t)$ and
defining $x_{f\text{ }}(t)\equiv x_{0}+\int\limits_{0}^{t}v_{f}(t^{\prime
})dt^{\prime },$ where $v_{f}$ is the time-dependent front velocity and $%
x_{0}$ is a constant. With the new set of independent variables $(\xi ,$ $%
B_{f})$ Eq. (\ref{GL}) becomes: 
\begin{equation}
\frac{v_{f}}{\Gamma }\left( -\frac{\partial \Psi }{\partial \xi }+\frac{%
\widetilde{B}}{d}\frac{\partial \Psi }{\partial B_{f}}\right) =D\frac{%
\partial ^{2}\Psi }{\partial \xi ^{2}}+\alpha _{0}\left( 1-\frac{%
B_{f}\left\{ x_{f}(t)\right\} }{B^{\ast }}-\frac{\widetilde{B}\xi }{dB^{\ast
}}\right) \Psi +\beta \Psi ^{2}-\gamma \Psi ^{3}.  \label{vfnonlin}
\end{equation}
One can solve this equation analytically provided the front width $\Delta
<<dB_{f}/\widetilde{B}$. In this case, the terms $(v_{f}/\Gamma )(\widetilde{%
B}/d)(\partial \Psi /\partial B_{f})$ and $-\alpha _{0}(\widetilde{B}\xi
/dB^{\ast })\Psi ,$ may be neglected. \cite{Keener} The solution of Eq. (\ref
{vfnonlin}) is then \cite{BenJacob}: 
\begin{equation}
\Psi =\Psi _{0}\left\{ 1+\exp \left( \frac{\xi }{\Delta }\right) \right\}
^{-1},  \label{Psigrowth}
\end{equation}
where 
\begin{equation}
\Delta ^{2}=\frac{\Delta _{0}^{2}}{2\mu (1-B_{f}/B^{\ast })+1+\sqrt{1+4\mu
(1-B_{f}/B^{\ast })}}  \label{delta}
\end{equation}
is the front width. The front velocity $v_{f}=dx_{f}/dt$ is: 
\begin{equation}
v_{f}=v_{0}\frac{6\mu (1-B_{f}/B^{\ast })+1+\sqrt{1+4\mu (1-B_{f}/B^{\ast })}%
}{\sqrt{2\mu (1-B_{f}/B^{\ast })+1+\sqrt{1+4\mu (1-B_{f}/B^{\ast })}}}.
\label{velocity}
\end{equation}
Here 
\begin{equation}
v_{0}=\Gamma \sqrt{D\beta ^{2}/4\gamma }=\Gamma \alpha _{0}d\sqrt{a_{D}}%
/2\mu ,\text{ }\Delta _{0}^{2}=4D\gamma /\beta ^{2}=4d^{2}a_{D},
\label{v0delta0}
\end{equation}
and $B_{f}\equiv B_{f}\left\{ x_{f}(t)\right\} $. The solid lines in Fig. 1
show $\Psi $ given in Eq. (\ref{Psigrowth})\ for different locations of the
front $x_{f}$, describing the propagation of the ordered phase.

We turn now to discuss the front velocity $v_{f},$ Eq. (\ref{velocity}), and
the front width $\Delta ,$ Eq. (\ref{delta}). We first note that $v_{f}$ and 
$\Delta $ do not depend {\em explicitly} on time or applied field but on $%
B_{f},$ the {\em local} induction at the front. Several important
conclusions may be drawn from these equations:

1) As $B_{f}$ approaches $B_{od}$\ the velocity is approaching zero (the $%
v_{f}(B_{f})$ dependence is described by the solid line in Fig. 2).

2) The motion of the front toward the sample edge is accompanied by an
increase of the induction $B_{f}$ at the front, resulting in a decrease of
the velocity with time.

3) The front width $\Delta $ decreases with the increase of $\beta $,
implying that for a 'stronger' first order transition the front is steeper.
Also from Eq. (\ref{delta}) it is obvious that the exchange coefficient $D$
causes the front to be smeared. In addition, increasing $D$ and/or the
damping coefficient $\Gamma $, results in an acceleration of the front
motion (see Eq. \ref{velocity}).

So far we have demonstrated dynamic coexistence of ordered and transient
disordered vortex phases, with a sharp interface between them, assuming
time-independent induction distribution with a constant gradient. In
high-temperature superconductors, however, the induction distribution varies
significantly with time due to flux creep. In addition, one may expect
different flux creep laws for the different vortex phases. As a result, the
nucleation and growth of the ordered vortex phase are manifested
experimentally by the appearance of a break in the induction profile and
movement of this break toward the sample edge. \cite{GillerPRL} The location
of the break is expected to coincide with the location of the moving front
of the order parameter.

To demonstrate this scenario we solved numerically the LK equation (\ref{GL}%
), allowing for flux creep. We assume a $\Psi $-dependent local current
density 
\begin{equation}
J(\Psi ,t)=J_{1}(t)\left( 1-\frac{\Psi }{\Psi _{0}}\right) +J_{2}(t)\frac{%
\Psi }{\Psi _{0}},  \label{JPsi}
\end{equation}
where $J_{1}(t)\ $and $J_{2}(t)$ are the current densities in the disordered
and in the ordered phases, respectively. \cite{Ej} Based on the experimental
observations \cite{GillerPRL} we further assume $%
J_{1}(t)=J_{01}(1+t/t_{1})^{-\alpha _{1}}$ and $%
J_{2}(t)=J_{02}(1+t/t_{2})^{-\alpha _{2}}$. The induction $B$ (and thus the
coefficient $\alpha $) may now be expressed in terms of the order parameter $%
\Psi (x)$ by using Maxwell equation: $B(x,t)=B_{a}-4\pi
/c\int\limits_{x}^{d}J(\Psi (y,t),t)dy$. As before, we assume an initial
disordered phase throughout the entire sample, $\Psi (t=0,x)=0$.

In order to solve equation (\ref{GL}) numerically we define dimensionless
parameters: $b=B/B^{\ast },$ $j=4\pi Jd/(cB^{\ast }),x^{\prime }=$ $x/d,$ $%
t^{\prime }=t\beta ^{2}\Gamma /\gamma ,$ $\Psi ^{\prime }=\Psi /\Psi
_{0}(B^{\ast \ast })=2\Psi \gamma /\beta .$ Eq. (\ref{GL}) then becomes: 
\begin{equation}
\frac{\partial \Psi ^{\prime }}{\partial t^{\prime }}=a_{D}\frac{d^{2}\Psi
^{\prime }}{dx^{\prime 2}}+\mu (1-b(x^{\prime }))\Psi ^{\prime }+\frac{1}{2}%
\Psi ^{\prime 2}-\frac{1}{4}\Psi ^{\prime 3}+f^{\prime }({\bf x}^{\prime
},t^{\prime }),  \label{LKdl}
\end{equation}
where $f^{\prime }=2f\gamma /(\beta \alpha _{0})$ is a dimensionless noise
which must be introduced in the numerical solution$.$

The values of the (dimensionless) parameters used in the numerical
calculations are based on experimental measurements. In particular, from the
fit of Eq. (\ref{velocity}) to the experimental data \cite{GillerPRL} of $%
v_{f}(B_{f})$ in Fig. 2 we estimate $\mu =1.5$, thus $B_{od}/B^{\ast
}=1.148, $ $B^{\ast \ast }/B^{\ast }=1.166.$ A value of $10^{-4}$ for $a_{D}=
$ $(2\mu v_{0}/(\Gamma \alpha _{0}d))^{2}$ (see Eq. \ref{v0delta0}) is
estimated from the experimental value $v_{0}\approx $ $20$ $\mu m/s$\
obtained from the same fit; a value of $\Lambda _{0}\sim \Gamma \alpha
_{0}\approx 10$ $s^{-1}$, is estimated from the time elapse between
switching-on of the external field and the appearance of a break in the
induction profile. A typical value of $d\approx 300$ $\mu m$ for the sample
half width was used. Based on the analysis of magnetic relaxation we take $%
\alpha _{1}=0.3,$ $\alpha _{2}=0.5$. In addition, a noise level of $f_{\max
}^{\prime }=10^{-4}$ is assumed. The system of equations completed by
boundary and initial conditions has been solved numerically utilizing the
Euler method. The unit space interval was divided into $200$ segments, and
time step of $2.5\cdot 10^{-3}$ (in dimensionless units) was used providing
a stability of the numerical procedure. The results for $B_{a}/B^{\ast }=1.1$
are shown in Fig. 3. The upper case shows the spatial dependence of the
order parameter at different times. The nucleation appears at the sample
center at $t^{^{\prime }}\sim 10$, forming a sharp front which propagates
toward the sample edge. \cite{YBCO} The lower case of Fig. 3 shows the time
evolution of the induction profiles during the nucleation and growth
processes. A sharp break in the profiles appears at the location of the
front of the order parameter after the nucleation is completed. As expected,
the break in the induction profile and the front of the order parameter move
together toward the sample edge. Note that a break in the induction profile
can be observed outside the region of phase metastability (i.e., for $%
B_{f}<B^{\ast })$.\ 

The theoretical predictions described above are confirmed experimentally in
BSCCO crystals. \cite{GillerPRL} In particular, breaks in the induction
profiles were recorded following a sudden application of external field of
intensity close to $B_{od}$. This break moves toward the sample edge at a
velocity which depends only on $B_{f}$, the value of the induction at the
break. Thus, the dependence of $v_{f}$ on $B_{f}$ is not affected by
magnetic relaxation. As shown in Fig. 2, the {\em analytical} curve (solid
line) of $v_{f}(B_{f})$, Eq. (\ref{velocity}), is in good agreement with the
experimental results (open symbols). Moreover, numerical results (solid
symbols in Fig. 2) for $v_{f}(B_{f})$\ for different applied fields also
show a good agreement with the analytical curve demonstrating that magnetic
relaxation does not affect the dependence of $v_{f}$ on $B_{f}$.

Finally, we note that two velocities govern the vortex dynamics in the
process of the phase transformation: The interface velocity $v_{f}$, and the
flux velocity, $v_{F}$, due to creep.\ The effect of the latter on the shape
of the interface must be taken into account in close vicinity of $B_{od}$\
where $v_{f}\longrightarrow 0$.

{\bf Acknowledgments}

This research was supported by The Israel Science Foundation founded by the
Israel Academy of Sciences and Humanities - Center of Excellence Program,
and by the Heinrich Hertz Minerva Center for High Temperature
Superconductivity. Y.Y. acknowledges support from the U.S.-Israel Binational
Science Foundation. D. G. acknowledges support from the Clore Foundation.

\begin{center}
\bigskip \newpage

{\LARGE Figure captions}
\end{center}

Figure 1. Nucleation and growth of the order parameter. The nucleation
process is demonstrated by the dashed curves calculated from Eq. (\ref{Psin}%
) for $A_{n}=A_{0}\delta _{n,0}$ at times $\Lambda _{0}t=8.12,$ $9.86$, $%
11.02$. The solid lines, describing the growth process, are calculated from
Eq. (\ref{Psigrowth}) at different locations\ $x_{f}/d=0.3,$ $0.5,$ $0.7,$ $%
0.9$.

Figure 2. Experimental data (open symbols) of $v_{f}(B_{f})$ for different
applied fields, taken from Ref. \cite{GillerPRL}, a fit (solid line) to Eq. 
\ref{velocity} and results of numerical solutions (solid symbols) for the
indicated applied fields . The fit was done with two fitting parameters $%
v_{0}$\ and $\mu ,$\ after $B_{od}$\ was estimated to be approximately 400
G, corresponding to an induction value where the velocity is going to zero.

Figure 3. Order parameter (upper case) and induction profiles (lower case)
for $j_{1}=1.7(1+t^{\prime })^{-0.3}$ and $j_{2}=1.5(1+t^{\prime })^{-0.5}$.
The profiles are shown for dimensionless times $t^{\prime }=1,$ $4,$ $9,$ $%
10,$ $11,$ $12,$ $14,$ $16,$ $18,$ $22,$ $27,$ $35,$ $44,$ $54,$ $65,$ $79,$ 
$99.$

\bigskip


\begin{references}
\bibitem{Bean}  As shown below, a field gradient is necessary for the
creation of front propagation of the equilibrium state.
\bibitem{Ej}  A more general approach to this problem should be based on
introduction of different E(j) characteristics for the different phases and
a solution of appropriate Maxwell equations. We solved our system using this
approach too, and obtained similar results.
\bibitem{YBCO}  The numerical solution does not exhibit breaks in the
profiles if the magnetic relaxation is too fast (e.g., $J\sim exp(-t/t_{0})$
while $t_{0}<<1/\Lambda _{0}$) or if the critical current is too low $%
(j_{01}\sim j_{02}\lesssim a_{D}/\mu )$. This is because the requirement for
a relatively large gradient, mandatory for the front development (Eq. \ref
{Condxsd}), is not satisfied in these cases. Thus, in $YBa_{2}Cu_{3}O_{7-%
\delta }$, for which $\widetilde{B}<<B_{od}\sim B^{\ast },$ an homogeneous
phase transformation is expected rather than front propagation.
\end{references}
\end{document}